# Machine learning – based diffractive imaging with subwavelength resolution

Abantika Ghosh[1#], Diane J. Roth[2#], Luke H. Nicholls[2#], William P. Wardley[2], Anatoly V. Zayats[2], and Viktor A. Podolskiy[1, *]

[1]*Department of Physics and Applied Physics, University of Massachusetts Lowell, Lowell, Massachusetts 01854, USA*

[2]*Department of Physics and London Centre for Nanotechnology, King's College London, Strand, London, WC2R 2LS, UK*

[#]*Equal contributions*

*Viktor_Podolskiy@uml.edu

**Far-field characterization of small objects is severely constrained by the diffraction limit. Existing tools achieving sub-diffraction resolution often utilize point-by-point image reconstruction via scanning or labelling. Here, we present a new imaging technique capable of fast and accurate characterization of two-dimensional structures with at least $\lambda/25$ resolution, based on a single far-field intensity measurement. Experimentally, we realized this technique resolving the smallest-available to us 180-nm-scale features with 532-nm laser light. A comprehensive analysis of machine learning algorithms was performed to gain insight into the learning process and to understand the flow of subwavelength information through the system. Image parameterization, suitable for diffractive configurations and highly tolerant to random noise was developed. The proposed technique can be applied to new characterization tools with high spatial resolution, fast data acquisition, and artificial intelligence, such as high-speed nanoscale metrology and quality control, and can be further developed to high-resolution spectroscopy.**

Optical characterization of an object implies transferring information about the shape and spectrum of this object from its location to the observer via electromagnetic waves and its resolution is determined by the diffraction limit[1-6]. Existing tools achieving sub-diffraction resolution rely on resonances to compensate or postpone exponential decay of evanescent radiation[7-12] or operate on extremely sparse, often luminescent, objects to achieve point-by-point image reconstruction[13-19], typically with multiple measurements per point. The environment separating the object and the detector determines the dispersion laws of the wave propagation and, therefore, plays a crucial role in limiting the quantity of information that can be relayed by optical means[4]. When the objects are large and well-isolated, optical systems operating in the ray-optics limit efficiently distribute their images across the image plane. This regime is well-suited for a multitude of rapidly emerging computer vision technologies[20-22] that generally rely on image segmentation followed by object detection and classification. Since edge detection is used throughout the computer vision workflow, sharp edges of well-separated objects are crucial for reliable operation of machine vision. Unfortunately, existing machine vision techniques are not readily applicable to highly diffractive configurations. Importantly, while computer vision tools increasingly use deep learning techniques with ever improving results, the exact information used by the algorithms to classify the images often remains unclear, making it almost impossible to correct the few, but important, misclassifications and predict the potential pitfalls. As the object size or separation are decreased, the deviations from ray optics due to diffraction become increasingly important, merging the images of multiple objects together and virtually eliminating the ability to resolve and identify small or closely spaced objects. Existing techniques to detect and classify small objects, including scanning optical microscopy, superlensing, structured illumination microscopy, fluorescent microsocopies, and sparsity-related super-resolution imaging typically rely on multiple measurements or reconstruct complex objects one point at a time[6-13]. However, when an object is positioned in the vicinity of a diffractive structure, light scattered by such a system carries substantial information about the object itself[23-25]. For one-dimensional objects and gratings, this information can be extracted analytically[23]. Similarly, the light scattered by a thin quasi-two-dimensional object positioned close to a two-dimensional diffractive structure carries the information about the object to the far field. However, since the increase in dimensionality necessarily yields exponential increase in complexity, the algorithms developed with line objects in mind cannot be directly applied to two-dimensional systems. Machine-learning tools, however, are implicitly robust in their ability to analyze complex patterns.

Here we demonstrate that artificial intelligence based on supervised learning can robustly deduce the structure of the diffractive objects with deep-subwavelength features based on properly parameterized far-field images. The developed technique employs the diffraction of light by a finite-size grating to boost the resolution of object characterization, resolving up to $\lambda_0/25$ features of the object, often with single, noise tolerant, intensity measurement. This novel diffractive imaging approach provides major speedup in characterization of small-scale features as compared with existing techniques, does not require scanning or multiple exposures, and can be further developed to high-resolution spectroscopy and new computer-assisted microscopy. We illustrate the capabilities of novel imaging paradigm by identifying the structure of a series of objects with subwavelength features coupled to finite diffraction gratings with lateral periods $\Lambda_x = 303 \pm 2\ nm, \Lambda_y = 335 \pm 2\ nm$, with the individual elliptical openings of short axis $r_x$ =83 ± 2 nm and long axis $r_y$ =90 ± 2 nm. To mimic the behavior of small objects of complex shape that block the propagation of light through some of the gratings openings, we used focused ion-beam-milling (see Methods) to fabricate a series of diffractive objects (summarized in Fig. 1 and in Methods). To avoid plasmonic enhancement effects, the structures were characterized using the light with a free- space wavelength of $\lambda_0 = 532\ nm$ (note that $r_{x,y} \simeq \lambda_0/5$), through the Fourier-optics-based diffractive imaging system (see Methods). The images measured by the CCD camera were then post-processed to remove the background noise and suppress the contributions of the main diffraction maxima. The resulting images are shown in Fig. 1.

The overall structure of these images is typical of a finite-size periodic grating. In particular, the two main maxima representing the zero- and first-order diffraction grating peaks, with their separation being proportional to inverse period of the grating (zero-order diffraction maximum is partially obscured by the lens glare). The positions and intensities of the auxiliary maxima represent the finer-scale structure of the grating, such as the number of openings, the relative transparencies of individual openings, etc.

In general, diffraction theory can be used to predict the pattern produced by a given structure. However, the inverse problem of identifying the structure of the grating based on the pattern at hand is a rather difficult one. It is clearly seen that while some diffractive images, for example, those of the objects P and D, produce signatures that are easily identifiable by the naked eye, the differences between the signatures produced by the majority of the samples are rather subtle. Artificial intelligence can accomplish the task of classification and identification of these complex image patterns.

In order to recognize the complex patterns that pertain to the particular objects, the AI system needs to be trained. The training set should mimic the experimental conditions and should convey the difference between experimental imperfections (i.e. abnormally sized or slightly moved opening) and an object completely blocking the particular set of openings. With this goal in mind, we digitally generated the training set of images. To produce the resulting library of diffractive signatures, the position and size of the holes in the theoretically-generated (phantom) gratings were randomly varied, with 2.5 nm variation in position and 10-nm variation in radius that mimic experimental conditions. For each object in the study 100 phantom gratings were generated. During the studies, this library was randomly sub-divided into the testing and training subsets.

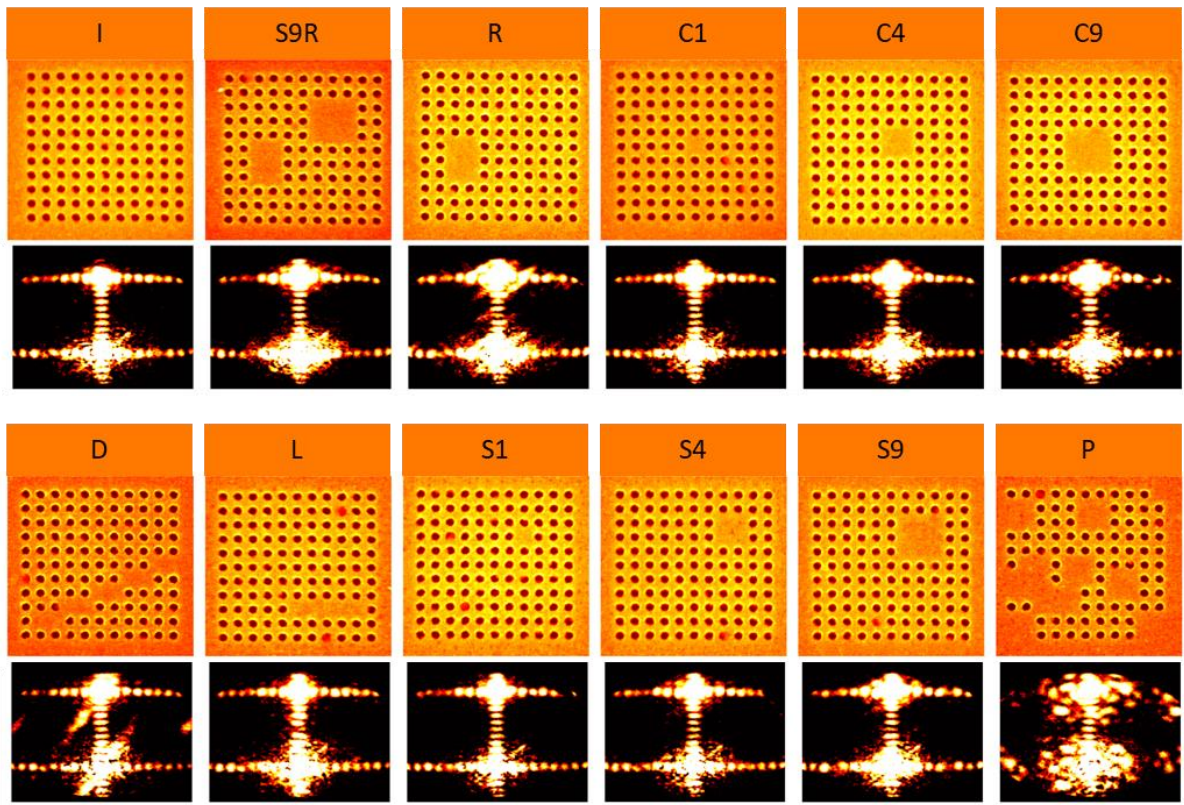

*Fig. 1. Specimens and their experimental diffractive signatures*. SEM images of the of the objects in the study (top sets) and post-processed diffraction imaging signatures of these objects (bottom sets). Nomenclature of the objects is given in Table 1.

To analyze the ultimate resolution limits of the diffractive imaging platform proposed here, similar image sets have been created assuming identical CCD parameters but longer operating wavelengths. In order to avoid resolution enhancement resulting from possible material resonances, we assumed that diffractive structures behave as perfectly opaque screens with perfectly transparent openings. As expected, the increase of the operating wavelength results in the expansion of the diffractive patterns, reducing the number of auxiliary maxima that fit within the numerical aperture of the system, and thus reducing the number of details that are available for characterizing the objects. Typical theoretical diffractive signatures of the same subset of phantom objects for different wavelengths of incoming light are illustrated in Fig. 2

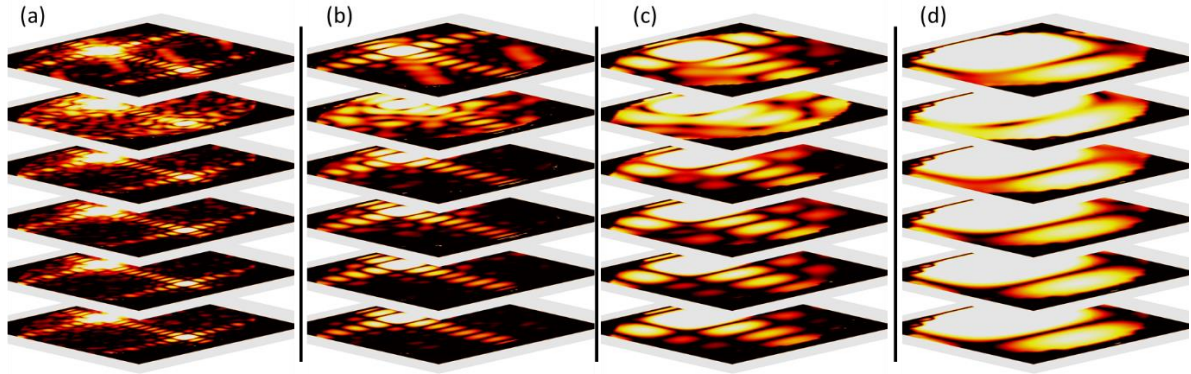

**Fig. 2. Diffractive signatures of numerical phantom objects.** *Diffractive signatures generated theoretically for free- space wavelength of (a) 532 nm, (b) 1 μm, (c) 2 μm, and (d) 4 μm. In each stack the images from bottom to the top represent objects I, S1, S4, R, P, and D.*

Once the training library of images is created, the problem of identifying the subwavelength objects is reduced to image classification, a three-stage supervised learning process that involves (i) the development of the approach that maps the image to its digital signature, (ii) training a computer classifier on the signatures of known objects, and finally (iii) the utilization of the trained classifier to identify (classify) unknown images based on their digital signatures. Note that once the classifier is trained, the actual recognition process is completed based on a single diffractive image that corresponds to the single experimental measurement.

The first stage of this process is the most crucial one since it is responsible for optimal encoding of optical information into digital form. Conventional techniques, built with computer vision in mind, call for image segmentation, edge identification, and analysis of edge distribution within the image[27-29]. However, while sharp edges do contain a significant portion of the information in ray-optics imaging, the onset of diffraction makes edges increasingly fuzzy. In addition, unavoidable CCD noise adds parasitic edges to the complex diffractive patterns. The combination of these phenomena renders the conventional computer vision tools almost unusable for diffractive imaging and new approaches are needed.

As seen in Fig. 2, the information carried by the diffractive optical system is encoded not in the edges of the diffraction maxima but in the distribution of their position, shape, and intensity. Such information can be readily extracted when the intensity of the diffractive image distribution is represented in the Bessel transform form[30],

$$I(k_r, \phi) \simeq \sum_{m,j} C_{mj} J_m \left( \alpha_{mj} \frac{k_r}{k_0} \right) \cos (m\phi) \qquad (1)$$

where $\phi$ is the polar angle, $\alpha_{mj}$ represents the $j$-th zero of Bessel function $J_m(x)$, the parameters $k_r$ and $k_0$ represent the radial component of the wavenumber and its maximal value, and the indices $m, j$ describe the behavior of the intensity in the angular and radial directions, respectively. Note that since both Bessel functions and cosines form orthogonal families of functions, Eq. (1) uniquely defines the values of the coefficients $C_{mj}$ independent of the number of terms in the sum.

In order to identify the subset of the components that carry the information about the most important features of the subwavelength objects, we utilize support vector machine (SVM)-based classifier [31-34] to analyze the library of diffractive images parameterized by a particular combination of $m, j$ pairs and analyze the accuracy of the resulting classifier as a function of the $m, j$ set. In this work, we limit ourselves to considering subsets of the coefficients where each of the indices $m, j$ is limited to the interval of a fixed length $l$.

Typical results of such a parameter sweep are shown in Fig. 3(a…d) where each point on the diagram represents recoveries based on the set of $l^2$ coefficients $C_{mj} … C_{m+l,j+l}$ for $l = 5$. To produce each individual point in the dataset, the numerically generated library of the diffractive signatures was separated into the training- and testing- subsets. The SVM was trained on the training subset, and its performance was validated by classifying the diffractive signatures of the objects from the testing subset. The original library was then split into different combination of training and testing subsets and the training/validation process was repeated.

It is seen that the majority of information is contained in lower-$m$ harmonics. At the same time, as the wavelength increases the useful information shifts towards smaller $j$ values, reflecting the slower oscillations of intensity in the radial direction with an increase of operating wavelength. Similar sweeps for larger values of parameter $l$, where each object is characterized by increasingly larger number of coefficients, suggest that the point of "maximum performance" approaches the origin ($m = 0, j = 1$). Combined, our analysis suggests that for best performance, the harmonics highlighted in Fig. 3(a…d) must be included in the final training and classification routines. The addition of harmonics that represent lower values of parameter $j$ (and thus classify slower oscillations) may improve the results. The analysis shows that the performance of the classifiers significantly degrades only when the operating wavelength reaches $4\mu m$, almost 20 times the period of the structure, and almost 80 times the radius of individual opening. To test the robustness of the developed platform, and to explore the potential effect of CCD hardware on the proposed algorithms random point noise was added to the simulated diffractive signatures [to simulate CCD noise], followed by Gaussian blur of the resulting images [to simulate CCD blooming] (see Methods) and the classification study was repeated. Our study indicates that the algorithm accuracy is virtually unaffected, even when ~50% of area of the simulated detector is affected by noise.

This analysis provides a valuable insight into the dynamics underlying the machine learning process, highlighting the relative importance of different components of the image for the resulting image classification, a process that is often hidden from view in conventional image recognition systems.

We further analyzed the accuracy the SVM achieves in identifying each individual object by averaging the data over multiple $\{m, j\}$ realizations. As expected, not all the objects are classified with the same accuracy. Generally, the more compact (smaller) an object is, the lower its classification accuracy.

We can therefore associate the size of the smallest object to be accurately classified with the resolution of the proposed diffractive imaging technique. Based on the analysis shown in Fig. 4, the resolution limit in our study is of the order of $\lambda_0/25$ at [corresponding to resolving S1 object at $\lambda_0 = 4\ \mu m$ with 50% accuracy]. However, this ideal resolution limit may be affected by various experimental factors. For example, the results presented here are affected by the presence of a lens glare, CCD saturation effects all of which can distort the diffractive information. The performance of the classifiers is likely to be also affected by the total number of elements in the finite diffraction grating, as well as by the number of objects that are being analyzed, with larger number of holes in the gratings or larger variety of the objects yielding smaller accuracy.

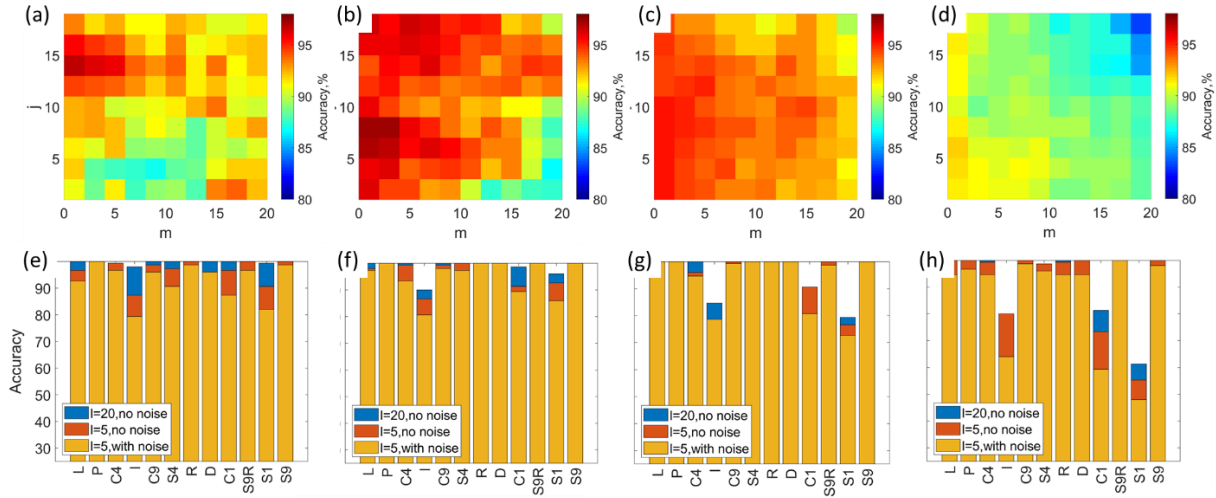

*Fig. 3. Theoretical performance of SVM classifiers. (a-d) Accuracy of the SVM classifiers trained on the subset of Bessel harmonics parameterized by the set of indices $\{m, j\}$ for $l = 5$ and (e-h) classification accuracy of the particular object for different operating wavelengths: (a,e) 532 nm, (b,f) 1 µm, (c,g) 2 µm, and (d,h) 4 µm.*

The size of the training set is an important parameter in the analysis of the performance of any AI-based system. Conventional ray-optics machine vision systems often require millions of training images to properly train a deep learning network. Recent theoretical studies of applications of convolutional neural networks to subwavelength imaging[35] confirm these trends, requiring $\sim 2 \times 10^4$ training sets and multiple "measurements" of both field amplitude and phase to resolve the dimensions and a separation between two 1D linear objects. Increase of dimensionality as well as limitation of intensity-based imaging tend to further increase the complexity of recovery algorithms and decrease the resulting resolution[4,25]

In contrast, the training process for the SVM-based 2D diffractive imaging is rather efficient, with only ~50 representations of each object being enough to ensure classification accuracy of above 80% for even the *worst-case*- S1 object for $\lambda_0 \leq 4\ \mu m$, based on a single far field diffractive intensity pattern (Fig. 4). Classification of larger objects is even more robust, with ~20 representations of each object being enough to achieve 95% accuracy for $\lambda_0 \leq 4\ \mu m$.

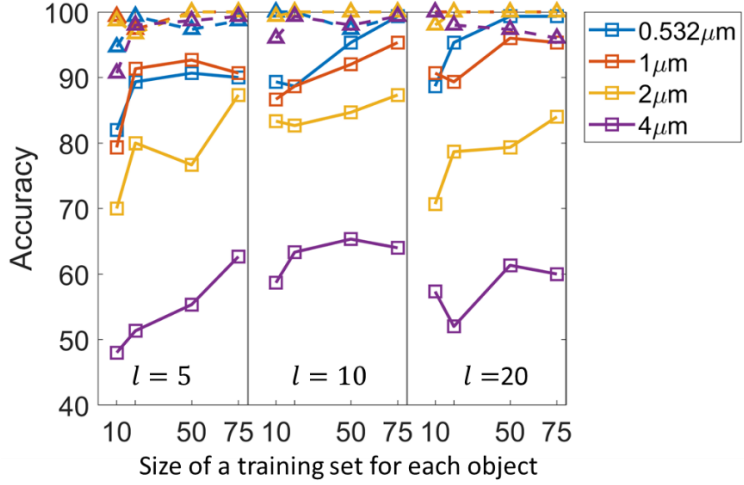

***Fig. 4. Convergence of the classifiers.*** *Classification accuracy as a function of the training set population for different values of the parameter $l$ that determines the number of $C_{mj}$ coefficients used in parameterization of each object: (solid lines) S1 object, which is the most challenging object to recover, and (dashed lines) S4 object.*

We now apply the developed formalism to classify the objects fabricated experimentally. In order to more comprehensively assess the perspectives of the machine-learning-based diffractive imaging, each sample was characterized with four different illumination directions, labeled here as Left, Right, Up, and Down, with data representing each direction analyzed independently. Statistical analysis of such recoveries is illustrated in Fig. 5, illustrating the robustness of the developed classifier that has been trained exclusively on theory-generated data that is capable of detecting each object (with smallest available to us dimension of 185 nm) with 532nm laser light. Note that while coherent laser light has been used in this work, our previous analysis[26] has indicated that gratings-assisted imaging works with incandescent white-light illumination. Similar to the theoretical studies reported above, not all the objects are classified with the same accuracy. However, in contrast to theory-based studies, the objects that exhibit the worst classification accuracy are not the smallest objects, a fact that likely reflects imperfections in experimental fabrication of the objects, as well as artifacts from lens glare and other experimental constraints.

Some of these constraints (lens glare or CCD blooming) are straightforward to incorporate in the developed formalism in practical settings where the experimental setup is used to characterize multiple similar objects. In these conditions the classifier can be trained on the diffractive signatures of known objects, therefore incorporating the systematic artifacts into the training process itself. We expect that in this scenario the performance of the final classifier would be comparable to the performance reported in the theoretical studies (Figs. 3,4).

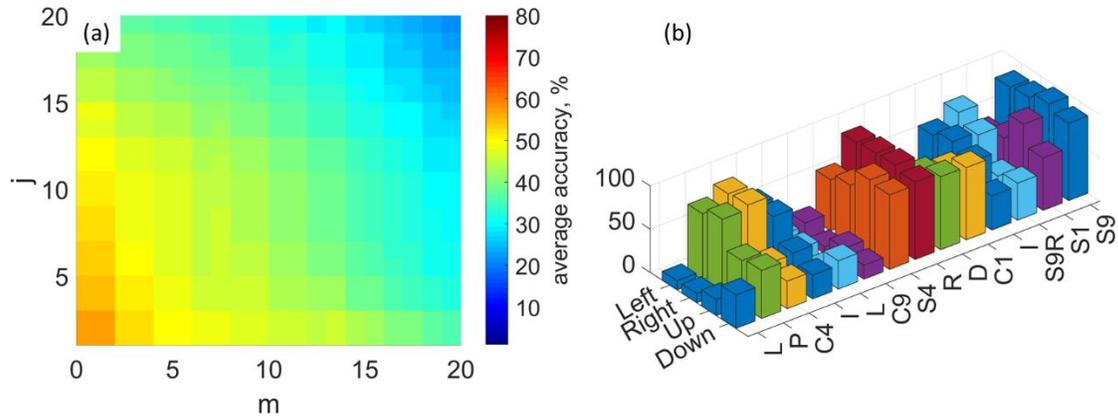

***Fig. 5. Characterization of the experimental objects***. *(a) The dependence of the accuracy of identification of the experimental objects in Fig. 1 for different combinations of $m, j$ parameters, averaged over $l \in [5,10,20]$, and different classifier settings. (b) Accuracy of characterization of the experimental objects as a function of orientation of the sample (Left, Right, Up, Down) for a subset of data shown in (a) resulting in recoveries of more than 5 samples.*

In summary, we have demonstrated the robust classification of subwavelength objects with diffractive imaging. Experimentally, smallest available to us objects of the order of $\lambda_0/3$ have been detected. Theoretical results suggest that the technique is highly tolerant to hardware noise and can be used to detect and classify smaller, at least $\sim\lambda_0/25$ objects with ~50% accuracy. Apart from demonstrating the new imaging approach, we have developed a robust algorithm for parameterization of diffractive images and identified the primary information flow channels used by the machine learning algorithms. As with any machine-learning techniques, the process of image recognition can be further optimized by providing training data that would more closely resemble experimental data with its systematic aberrations. In this sense, classification performance can be improved if realistic imaging is trained on experimental data, not on idealized theoretical predictions. All in all, the proposed technique opens the door for robust classification and characterization of objects with subwavelength structure, including fast and robust quality control in nanofabrication, and optical analysis of nano-structural fingerprints of complex objects. The same approach can be used to analyze, in transmission geometry, the structure and the spectrum of small objects positioned above the finite diffraction gratings. In this scenario the object blocks a number of the openings of the diffraction grating, and thus modifies the transmission through the structure. A properly trained classifier is then used to identify the subset of the blocked holes and thus, to recover the structural information about the object. Repeating the same process for different incident wavelength yields spectral, in addition to structural, information about the object.

## Acknowledgements
The research was partially supported by the Army Research Office (US) Grant #W911NF-16-1-0261, EPSRC (UK), and ERC iCOMM project (789340).

## Author contributions
VAP and AVZ designed the theoretical and experimental studies and supervised the research. WPW and DJR fabricated the samples. DJR and LHN designed and performed the experiment. AG generated numerical phantoms, trained the classifiers and analyzed their performance. All authors contributed to manuscript writing.

**Competing Interests**

The authors declare no competing interests.

## METHODS

**Sample fabrication**

The 2-dimensional gratings were fabricated in a 100 nm thick gold film using focused ion beam milling. The gold film was deposited on a glass coverslip covered with a 15 nm tantalum pentoxide ($Ta_2O_5$) adhesion layer using a DC magnetron sputtering. The ideal structure, without defects, consists of an 11x11 array of elliptical holes with a 165 ± 2 nm short axis and 180 ± 2 nm long axis; two ideal structures have been fabricated. The lattice periods are 303 ± 2 nm and 335 ± 2 nm in the short and long axis directions, respectively. Various geometrical defects were introduced in the fabrication of the other gratings by omitting holes, producing variations in the Fourier diffraction patterns experimentally and theoretically observed. Defects included single missing holes and square patterns of 2x2 or 3x3 missing holes, located either in the centre of the array or at a random position in the structure; a 2x3 rectangle; the combination of a 3x3 square and 2x3 rectangle; a diagonal line defect; a 1x5 straight line (two structures have been fabricated); as well as a random pattern of missing holes. Each of the 12 types of objects used in the study is assigned a unique legend consisting of a (set of) letters and numbers (Table S1).

Table S1. Nomenclature of the studied objects shown in Fig. 1.

| Label | Object Description | Label | Object Description |
|---|---|---|---|
| I | Ideal grating, no intentional defects | S9R | Grating with a 9-hole square and a 6-hole rectangular area blocked |
| R | Grating with a 6-hole rectangular area blocked | C1 | Grating with a center hole blocked |
| C4 | Grating with a 4-hole square area in the center blocked | C9 | Grating with a 9-hole square area in the center blocked |
| D | Grating with 8-hole oblique line blocked | L | Grating with a 5-hole horizontal line blocked |
| S1 | Grating with a single off-center hole blocked | S4 | Grating with a 4-hole off-center square area blocked |
| S9 | Grating with a 9-hole off-center square are blocked | P | Grating with a quasi-random pattern blocked |

**Optical measurements**

Optical measurements were performed using a similar experimental setup (Fig. S1) as the one described in Ref. [26]. The structures were illuminated by a quasi plane wave, generated by focusing a 532 nm CW laser beam onto the back focal plane of a 40 x objective (0.95 NA) incident on the metal film. The angle of incidence on the sample was controlled by displacing the focal spot onto the back focal plane of the illumination objective. The scattered light from the structures was collected in transmission through the substrate by an oil immersion 100X objective (NA= 1.49). The back focal plane of the detection objective (Fourier plane) was then imaged onto an imaging spectrometer using a set of relay lenses.

For each structure, a measurement at normal incidence from the sample was taken, along with a set of measurements at an angle of incidence of 50° for four cardinal orientations of the grating (needs a figure). The power of the laser was set to 150 µW and two sets of measurements were then recorded for exposure times of 10 ms and 40 ms, in order to collect more intensity in the higher diffraction orders for analysis. Background images on the gold film were also recorded.

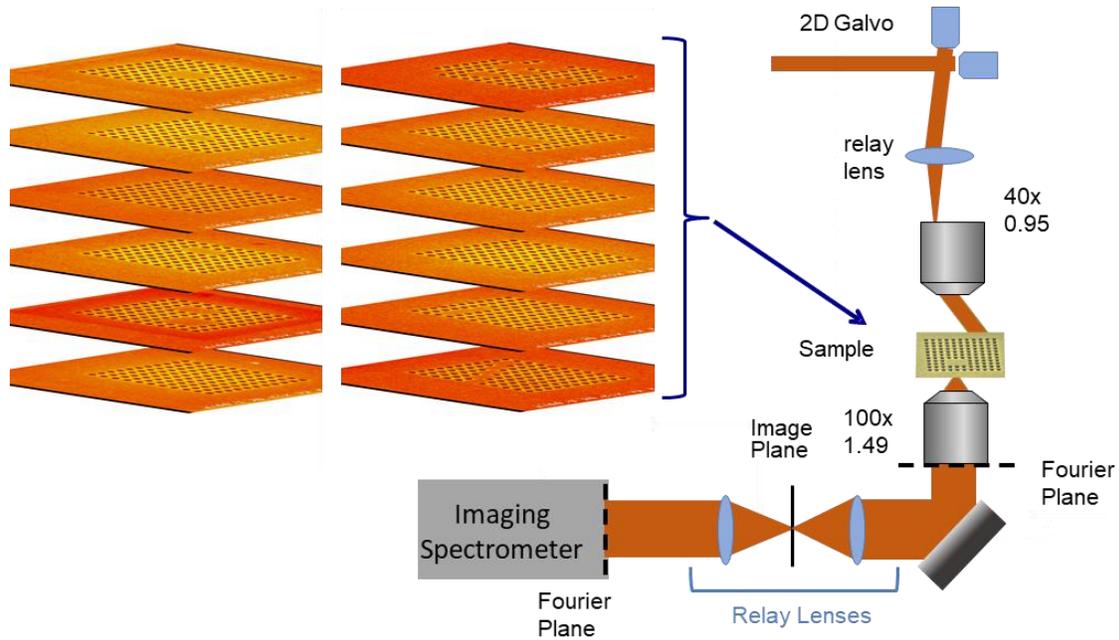

*Fig. S1. Optical setup for performing diffractive imaging. Inset shows the SEM images of the objects used in the study.*

**Theoretical generation of library of images**

Far field (Fraunhofer) approximation is used to generate the Fourier signatures of different samples in our theoretical studies. Numerically, we begin with computer-generated binary mask representing the geometry of the particular grating, with openings randomly displaced from the "ideal" periodic grating by at most 2.5 nm, and with randomly generated opening radius of $80 \pm 10 nm$. We assume that the openings of the grating are fully transparent to the (normally incident) monochromatic plane wave while the space between the openings fully blocks the incoming radiation. Therefore, the spatial distribution of the electromagnetic field just behind the grating is proportional to the binary mask of the grating itself.

This spatial profile is then Fourier-transformed, and the bandwidth of the resulting Fourier representation is cut to mimic the numerical aperture of the optical setup used in the experiments. To mimic the aberration of the experimental setup, we followed the coordinate transformation $\{k_x, k_y\} \rightarrow \{k_x, k_y[1 - \alpha\, k_x^2]\}$, with the value of the parameter $\alpha$ based on experimental images. The slight ellipticity of the holes in the experiment has been neglected as the deviations from the circular holes ($\pm 15$ nm) are beyond the resolution of the proposed set-up.

**Image post-processing**

Prior to machine learning analysis, each CCD image was post-processed according to the following algorithm. First, the background pattern (representing transmission through smooth gold film) was subtracted. Next, the CCD noise and the saturated signals were discarded (by imposing lower and upper cut-off values). CCD signals representing CCD space outside the numerical aperture of the imaging signal was discarded as well. Finally, the intensity was converted to the log scale to enhance the diffractive signals. Theoretically-produced intensity distributions were post-processed in similar fashion. In all analyses, only the portion of the image representing $0 \leq k \leq k_0, -\frac{\pi}{2} \leq \phi \leq \frac{\pi}{2}$ [see Eq.(1) and Fig.S2] was used.

**Setup of the support vector machine**

Support vector machines (SVM) implementation outlined in Ref.[31-33] was used in this work. To understand and optimize the information flow through the system, we have analyzed the recovery accuracy of multiple SVMs, with linear, polynomial, as well as Gaussian kernels, and with different multiclass classification combinations. Our analysis suggests that linear kernel with the multiclass classifier that relies on the array of one vs. all binary SVM sub-classifiers performs the best for the diffractive classification problem.

**Supporting Information.**

**Adding noise to the simulations**

To assess the effect of the random noise (for example, generated by the CCD) on the performance of diffractive imaging, the training and recovery procedures were repeated on the noise-affected theoretical images. Each noise-affected image was formed by starting with its no-noise "baseline" counterpart and adding a set of Gaussian noise spikes at random locations of the image. The level of noise is parameterized by the fraction of the total area occupied by the noise spikes. Fig.S1 illustrates this process.

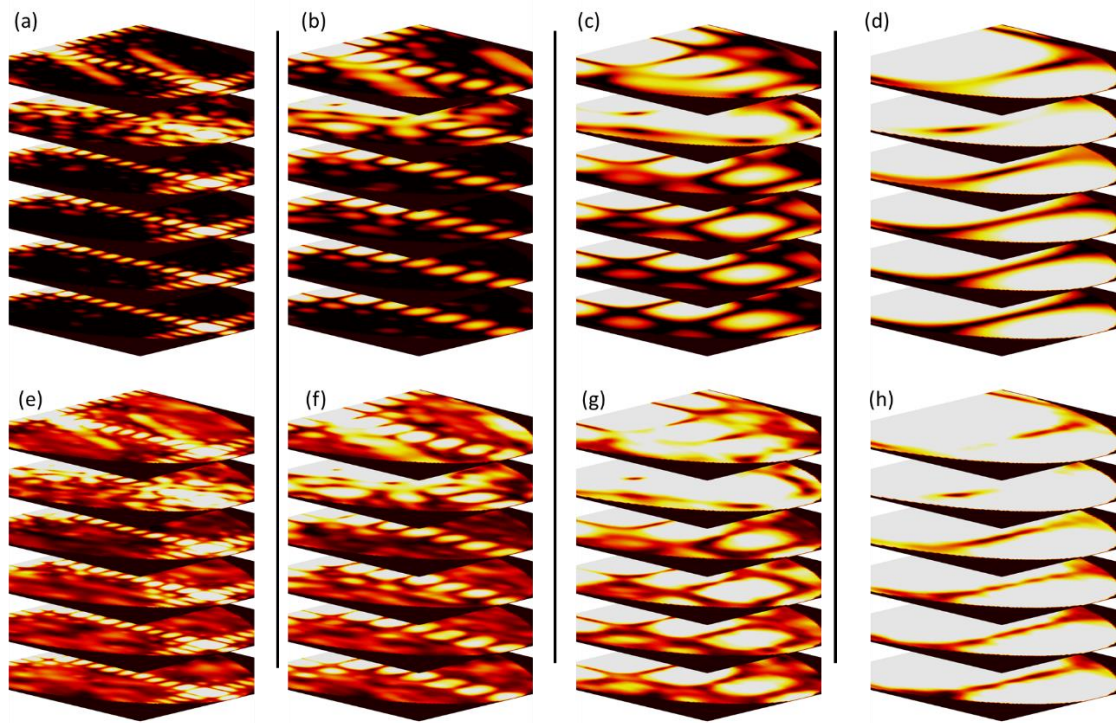

*Fig. S2. Diffractive signatures generated theoretically for a free space wavelength of 532 nm (a,e), 1 $\mu m$ (b,f), 2 $\mu m$ (c,g), and 4 $\mu m$ (d,h) with no noise (a-d) and with 50% added noise (e-h). In each stack the images represent (from bottom to top) represent objects I, S1, S4, R, P, and D, respectively.*